\documentclass{article}

\usepackage{microtype}
\usepackage{graphicx}
\usepackage{subcaption}
\usepackage{booktabs} 

\usepackage{hyperref}

\usepackage[preprint]{icml2026}

\usepackage{amsmath}
\usepackage{amssymb}
\usepackage{mathtools}
\usepackage{amsthm}

\usepackage[capitalize,noabbrev]{cleveref}

\theoremstyle{plain}

\theoremstyle{definition}

\theoremstyle{remark}

\usepackage[textsize=tiny]{todonotes}

\usepackage{caption}
\usepackage{url}
\usepackage{xspace}
\usepackage{multirow}
\usepackage{amsfonts}       
\usepackage{nicefrac}       
\usepackage{microtype}      
\usepackage[table,xcdraw]{xcolor}
\usepackage{tabularray}
\usepackage{enumitem}
\usepackage{pifont}
\usepackage{wrapfig}

\newcommand{\blackcircled}[1]{%
  \tikz[baseline=(char.base)]{
    \node[shape=circle, fill=black, text=white, inner sep=1pt, font=\sffamily\bfseries\scriptsize] (char) {#1};}}
\newcommand{\rfm}{\textit{RefactoringMiner}\xspace}
\newcommand{\astm}{AST-Based Refactoring Verification\xspace}
\newcommand{\puc}{\textit{PurityChecker}\xspace}
\newcommand{\tool}{\textit{SWE-Refactor}\xspace}

\newcommand{\numproj}{18\xspace}
\newcommand{\numref}{1,099\xspace}
\newcommand{\peter}[1]{\textcolor{blue}{{\it [peter says: #1]}}}

\usepackage[most]{tcolorbox}
\usepackage{verbatim}

\tcbset{
  promptbox/.style={
    colback=gray!5,
    colframe=black!60,
    fonttitle=\bfseries,
    top=1mm, bottom=1mm, left=2mm, right=2mm,
    boxrule=0.4pt,
    sharp corners,
  }
}

\newcommand{\cmark}{\ding{51}}  
\newcommand{\xmark}{\ding{55}}  

\icmltitlerunning{SWE-Refactor: A Repository-Level Benchmark for Real-World LLM-Based Code Refactoring}

\begin{document}

\twocolumn[
  \icmltitle{SWE-Refactor: A Repository-Level Benchmark for Real-World LLM-Based Code Refactoring}



  \icmlsetsymbol{equal}{*}

  \begin{icmlauthorlist}
    \icmlauthor{Yisen Xu}{yyy,comp}
    \icmlauthor{Jinqiu Yang}{comp}
    \icmlauthor{Tse-Hsun (Peter) Chen}{yyy}
  \end{icmlauthorlist}

  \icmlaffiliation{yyy}{SPEAR Lab, Concordia University, Montreal, Canada}
  \icmlaffiliation{comp}{O-RISA Lab, Concordia University, Montreal, Canada}

  \icmlcorrespondingauthor{Tse-Hsun (Peter) Chen}{peterc@encs.concordia.ca}

  \icmlkeywords{Machine Learning, ICML}

  \vskip 0.3in
]



\printAffiliationsAndNotice{}  

\begin{abstract}
  Large Language Models (LLMs) have recently attracted wide interest for tackling software engineering tasks. In contrast to code generation, refactoring demands precise, semantics-preserving edits that improve program structure, which also makes automated evaluation challenging. However, existing refactoring benchmarks commonly suffer from three shortcomings: limited coverage of refactoring scenarios, the inclusion of instances that mix refactoring with unrelated changes, and insufficient repository-level context for realistic assessment. To mitigate these issues, we introduce SWE-Refactor, a new benchmark for LLM-based code refactoring. SWE-Refactor comprises 1,099 developer-written, behavior-preserving refactorings mined from 18 Java projects, including 922 atomic and 177 compound instances. Each instance is validated via compilation, test execution, and automated refactoring detection tools to ensure correctness. We evaluate nine widely used LLMs on SWE-Refactor, covering models such as GPT-4o-mini, DeepSeek-V3, and CodeLLaMa, to provide representative reference results. Our results show that complex and compound refactorings remain the primary source of failures; notably, an OpenAI Codex agent achieves only 39.4\% success on compound instances. We release SWE-Refactor and all evaluation results to facilitate future research on LLM-based code refactoring.
\end{abstract}

\section{Introduction}
\label{sec:introduction}





In software engineering, code refactoring is a process of improving the structure of existing code without changing its behavior~\citep{DBLP:books/daglib/0019908}. This practice is essential for maintaining software systems by improving code quality, enhancing reusability, and ensuring adaptability to changing requirements  \citep{murphy2011we}. Unlike coding, code refactoring typically involves analyzing existing code to identify code segments for improvement, understanding its structure and dependencies, and making precise changes without altering its behavior. For example, a common refactoring operation is \textit{Extract Method}~\citep{DBLP:books/daglib/0019908, murphy2011we, tsantalis2020refactoringminer}, where a developer identifies a portion of a long method that can operate independently and extracts it into a separate method, making the original method shorter, more readable, and reusable.

Among these tasks, code generation has attracted significant attention \citep{lin2024soen101codegenerationemulating, DBLP:journals/corr/abs-2406-00515, DBLP:journals/corr/abs-2404-02183}, where LLMs generate code from natural language descriptions or specifications. Refactoring, however, is a common software maintenance activity \citep{DBLP:conf/sigsoft/IversNOSTK22} that poses distinct challenges. Unlike code generation, refactoring stresses LLMs’ capabilities for repository-level reasoning and for preserving behavior. LLMs must reason about existing code and dependencies, and then apply precise, semantics-preserving transformations. Refactoring further challenges agentic frameworks, since it often requires iterative planning, coordinated code edits, and repeated verification to ensure behavior is preserved.
Moreover, evaluating refactoring capabilities requires realistic settings and codebases, since real-world code introduces complex design patterns, dependency chains, and language features that are rarely captured in synthetic examples.


To assist with these challenges, mainstream integrated development environments (IDEs) such as IntelliJ IDEA \citep{Intellij_IDEA}, PyCharm \citep{PyCharm}, and Eclipse \citep{Eclipse} have introduced semi-automated refactoring tools. These tools can help perform low-level code changes but still rely heavily on developers to understand the code and make key decisions. To further reduce manual effort and enhance automation, recent studies have investigated the use of LLMs for code refactoring tasks \citep{pomiannext, shirafuji2023refactoring, white2024chatgpt, DBLP:journals/corr/abs-2503-14340}, and several benchmarks have been proposed to evaluate model performance. 
However, as summarized in Table~\ref{table:statistic}, these benchmarks often have one or more of these four key limitations listed below.

\begin{table*}
\centering
\caption{The comparison between existing benchmarks and \tool. \textbf{Compound refactoring} means there can be multiple code transformations.
\textbf{Pure refactoring} indicates commits without unrelated changes. 
\textbf{Developer-written GT} refers to the ground truth refactored code being written by \textit{original project developers}. 
\textbf{Test availability} shows whether test cases are provided to verify correctness. 
\textbf{Automated construction} indicates whether the benchmark was built entirely via an automated pipeline.}
\label{table:statistic}
\renewcommand{\arraystretch}{1.5}
\resizebox{\textwidth}{!}{
\begin{tabular}{llllllll}
\toprule
\multirow{2}{*}{\textbf{Benchmark}} & \multicolumn{2}{c}{\textbf{Code Distribution}} & \textbf{\textbf{Compound }} & \textbf{Pure} & \multicolumn{1}{c}{\textbf{Developer-}} & \textbf{Test~} & \multicolumn{1}{c}{\textbf{Automated}} \\ 
\cline{2-3}
 & \textbf{\# Repo} & \textbf{\# Sample} & \textbf{Refactoring} & \textbf{Refactoring} & \textbf{\textbf{Written GT}} & \textbf{\textbf{Availability}} & \textbf{Construction} \\ 
\hline
{ref-Dataset}~\citep{DBLP:journals/ase/LiuJZNLL25} & 20 & 180 & \xmark & \cmark & \cmark & \xmark & \xmark \\ 

{community corpus}~\citep{pomiannext} & 5 & 122 & \xmark & \xmark & \cmark & \xmark & \xmark \\ 

{extended corpus}~\citep{pomiannext} & 12 & 1,752 & \xmark & \xmark & \cmark & \xmark & \cmark \\ 

{RefactorBench}~\citep{RefactorBench} & 9 & 100 & \xmark & \xmark &\xmark & \cmark & \xmark \\ 

\tool & \numproj & \numref &\cmark & \cmark & \cmark & \cmark & \cmark \\
\bottomrule
\end{tabular}
}
 \vspace{-3mm}
\end{table*}

\blackcircled{1} \textbf{Consider Only Atomic Refactoring Types}.
Existing refactoring benchmarks often focus on a limited set of \textbf{\textit{atomic refactoring types (i.e., a single code transformation)}}.
Figure~\ref{fig:overview} shows an example where an \textit{Extract Method} appears \textit{\textbf{as part of}} a \textit{\textbf{compound refactoring (i.e., multiple code transformations)}}. 
As shown in Table~\ref{table:statistic}, existing benchmarks predominantly cover atomic refactorings. However, compound refactorings are also common in real development, where a single maintenance change often requires multiple coordinated transformations. Compound refactorings can  be more challenging for LLMs, because models must maintain cross-step and cross-file consistency (e.g., updating call sites, signatures, and dependencies) while preserving behavior.

\blackcircled{2} \textbf{Lack of Automated Construction.} 
Many existing benchmarks are not automatically constructed, requiring manual effort in various stages such as preparing pre-refactoring code or writing ground truth and test cases.
Specifically, \textit{ref-Dataset}~\citep{DBLP:journals/ase/LiuJZNLL25} manually reverts code changes to reconstruct pre-refactoring code, which is both time-consuming and error-prone. 
\textit{RefactorBench} manually constructs, with the help of LLM, both the ground truth refactored code and the corresponding test cases. 
These manual steps make the benchmarks difficult to scale and maintain. 
Some changes even go beyond refactoring, such as modifying repository logic, which shifts the focus away from behavior-preserving code refactorings.

\blackcircled{3} 
\textbf{Insufficient Support for Repository-Level Analysis and Automated Verification}. Existing refactoring benchmarks are not designed to evaluate LLM's capability in repository-level tasks. They typically include only basic elements such as task descriptions, code before and after refactoring, and lack the additional repository-level information (e.g., method callers and callees, class hierarchies, and inheritance relationships) required for more advanced refactoring or repository-level analyses. 
Moreover, most benchmarks do not provide tests for automated verification. 
Among all existing benchmarks, only \textit{RefactorBench} \citep{RefactorBench} includes associated tests. 
\blackcircled{4} \textbf{Noisy Benchmark Data.} Existing refactoring benchmarks often contain code changes that are not purely refactoring. This occurs because refactoring activities are mostly driven by changes in requirements (such as new features and bug fixes), and less driven by solely code smell resolution \citep{DBLP:conf/sigsoft/SilvaTV16}. 
However, impure changes make it hard to determine whether the LLM-generated code aligns with the intended refactoring. If the reference solution contains both refactorings and other functional changes, it becomes unclear which types of changes the model is expected to generate. This ambiguity reduces the effectiveness of benchmarks for evaluating code refactoring. 
As shown in Table \ref{table:statistic}, among all existing benchmarks, only  \textit{ref-Dataset}~\citep{DBLP:journals/ase/LiuJZNLL25} contains pure refactorings, where the authors manually removed the refactoring from the modified code to recreate the original version. This method works for simple refactorings, such as \textit{Rename Method}, but is hard to apply to more complex cases that involve multiple files, like \textit{Move Method}, due to manual overheads. 

Existing software engineering benchmarks also suffer from a significant imbalance in programming languages. A recent study by ~\citet{DBLP:conf/kbse/CaoCWC024} shows that 95.6\% of the latest benchmarks are built exclusively on Python (e.g., SWE-bench~\citep{DBLP:conf/iclr/JimenezYWYPPN24}, HumanEval~\citep{DBLP:journals/corr/abs-2107-03374}, MBPP~\citep{DBLP:journals/corr/abs-2108-07732}, and RefactorBench~\citep{RefactorBench}), limiting the diversity and representativeness of evaluation. 
To bridge this gap and address the above-mentioned challenges, we introduce \tool, a benchmark for evaluating LLMs’ code refactoring capabilities on Java projects. Java is one of the most widely used programming languages in the world, ranking among the top in both the TIOBE index~\citep{tiobe2025} and the Stack Overflow developer survey~\citep{stackoverflow2024}. 
Java's statically typed and syntactically structured grammar also results in well-defined refactoring patterns, allowing for more precise and accurate refactoring benchmarking. By focusing on Java, \tool broadens evaluation beyond the current Python-centric landscape and reflects the languages used in large enterprise and open-source systems. 

\tool consists of \numref pure refactorings extracted from \numproj widely used Java projects, complementing existing benchmarks (e.g., \textit{RefactorBench}) that predominantly focus on Python. \blackcircled{1} \textbf{In addition to atomic, it also covers compound refactoring types}, including three atomic types—\textit{Extract Method}, \textit{Move Method}, and \textit{Inline Method}—as well as three compound types—\textit{Extract and Move Method}, \textit{Move and Inline Method}, and \textit{Move and Rename Method}.  \blackcircled{2} \textbf{\tool eliminates noise and includes only pure refactorings}. To ensure the purity of refactoring, we use abstract syntax tree (AST)-based refactoring detection tools that are shown to have great precision (98\%) and recall (91\%)~\citep{tsantalis2018accurate, tsantalis2020refactoringminer, nouri2023puritychecker} 
to extract and select only pure refactoring from a large number of real-world refactoring code commits. 
To validate that these detectors work reliably in \tool, we further manually inspect 200 randomly sampled instances and confirm that they are behavior-preserving refactorings without unrelated changes. 
\blackcircled{3} \textbf{\tool provides comprehensive repository-level information}. 
In addition to the basic information (code before refactoring, developer-written refactored code, and refactoring type), \tool provides rich repository-level and structure information, including project structure, class body, caller and callee of method, build configuration details, and test coverage information. \blackcircled{4} \textbf{\tool ensures automated and reproducible data collection}.
\tool fully automates the extraction of pure refactoring data from real-world projects, avoiding the need for manual annotation or LLM-generated code. All ground-truth refactored code is directly derived from project repositories. This ensures scalability and future benchmark expansion. 
\blackcircled{5} \textbf{High quality and executable refactoring.} \tool extracts developer-written refactorings from real-world projects with diverse application domains, allowing it to better reflect the capabilities of LLMs in realistic software engineering scenarios. 

To ensure the reliability of the benchmark, we perform multi-stage verification: (i) AST-based static analysis to confirm that each commit contains only the targeted refactoring type and no unrelated code changes, (ii) compilation and execution of the full test suite to confirm behavioral equivalence, and (iii) manual checks on a subset of instances to prevent false positives from automated tools.
We retain only those refactorings that pass all verification steps, ensuring that \tool contains high-quality, executable, and behavior-preserving examples.
Appendix~\ref{appendix:dataset} shows the details on the project selection and refactoring distributions. 


We evaluate 9 widely used LLMs: GPT-4o-mini \citep{DBLP:journals/corr/abs-2303-08774}, GPT-3.5 \citep{GPT3.5}, DeepSeek V3 \citep{DBLP:journals/corr/abs-2412-19437}, Qwen2.5 Coder (14B, 7B) \citep{DBLP:journals/corr/abs-2409-12186}, DeepSeek Coder (16B, 6.7B) \citep{DBLP:journals/corr/abs-2401-14196}, and CodeLLaMA (13B, 7B)\citep{DBLP:journals/corr/abs-2308-12950}, on \tool. We evaluate the refactored code along two dimensions: functional correctness and human-likeness. For functional correctness, we assess the code using 1) compilation success and project test pass rate, and 2) \astm, which verifies that the expected refactoring has indeed occurred in the modified code. For human-likeness, we use \textit{CodeBLEU} \citep{ren2020codebleu} to measure similarity to developer-written refactorings. Since CodeBLEU mainly captures surface-level similarity, we use it only for human-likeness. 

Overall, larger general-purpose models tend to perform better than smaller open-source models on this benchmark. Performance also varies across refactoring types: models do well on some atomic refactorings but remain challenged on complex and compound refactorings. Notably, OpenAI Codex (\texttt{GPT-5.1-Codex}) \cite{gpt-5-codex}, an agent designed for real-world coding tasks, achieves only 39.4\% success rate on compound refactorings.


Overall, our contributions in this work are threefold:
\begin{itemize}[topsep=0pt,itemsep=2pt,parsep=0pt]
\item 
We introduce \tool, a benchmark constructed from developer-written commits that contain only refactorings and no other functionality changes. It is designed to evaluate LLM's capabilities on both atomic and compound refactoring tasks.
\item We design a fully automated four-step pipeline to construct \tool, which extracts real refactorings, filters out impure ones, collects relevant structural information, and verifies functional correctness through compilation and test execution.
\item We conduct an extensive evaluation of 9 popular LLMs on \tool and perform a fine-grained analysis of their performance across different refactoring types, highlighting their strengths and limitations.

\item We release \tool, the evaluation code, and all results; details and access instructions are provided in Appendix~\ref{appendix:dataset-hosting}.

\end{itemize}

\section{Related Work}
\label{sec:related_work}

\textbf{Refactoring Benchmarks.}
\textit{RefactorBench}~\citep{RefactorBench} is a Python-based benchmark for evaluating the effectiveness of LLMs in code refactoring. 
Unlike \tool that leverages developer-written refactorings mined from real commits, RefactorBench relies on LLMs to 
identify refactoring opportunities, which can introduce model-specific biases into the benchmark. Moreover, \tool captures the complex real-world software design, including overridden methods, generics, exception handling, and inheritance that are often missing in synthetic data. 

RefactorBench's ground truth solutions are also manually written by the authors, who may not have in-depth knowledge of the project. 
\textit{ref-Dataset} \citep{DBLP:journals/ase/LiuJZNLL25} includes 100 pure atomic refactorings from real Java projects. The \textit{community corpus} provides 122 Extract Method refactorings from five older Java projects. The \textit{extended corpus} \citep{pomiannext} expands this to 1,752 Extract Method instances. However, each of the benchmarks has its own limitation, as shown in Table~\ref{table:statistic}. Our benchmark, \tool, is automatically built from \numproj modern Java projects, covering both atomic and compound refactorings. All ground truth refactored code and test cases are written by the original project developers. The benchmark supports automated evaluation and ensures both structural and behavioral correctness through compilation and full test verification.

\textbf{LLMs-based Code Refactoring.}
Recent works have explored various techniques to enhance LLM performance in refactoring tasks, including prompt clarity~\citep{alomar2024refactor}, structured prompting~\citep{white2024chatgpt}, and few-shot learning~\citep{shirafuji2023refactoring}. Hybrid approaches that combine LLMs with rule-based systems have also shown improved results~\citep{zhang2024refactoring}. Several works directly prompt models like GPT-4 to perform refactorings~\citep{depalma2024exploring, poldrack2023ai}, confirming the feasibility of using LLMs for this task.
In addition, practical tools such as \textit{EM-Assist}~\citep{pomiannext} and the Context-Enhanced Framework~\citep{gao2024context} demonstrate how LLMs can be integrated into automated refactoring workflows.
A recent study~\cite{DBLP:journals/corr/abs-2503-14340} explored using agents for automated refactoring. However, the paper did not provide the benchmark or source code for replication. 
Our benchmark can serve as a basis for future work in this area by providing a standardized and real-world dataset to evaluate and compare refactoring capabilities of LLMs across both atomic and compound transformations. 

\section{SWE-Refactor}
\label{sec:benchmark}


\subsection{Overview}
\begin{figure*}[!tb]
  \includegraphics[width=0.7\textwidth]{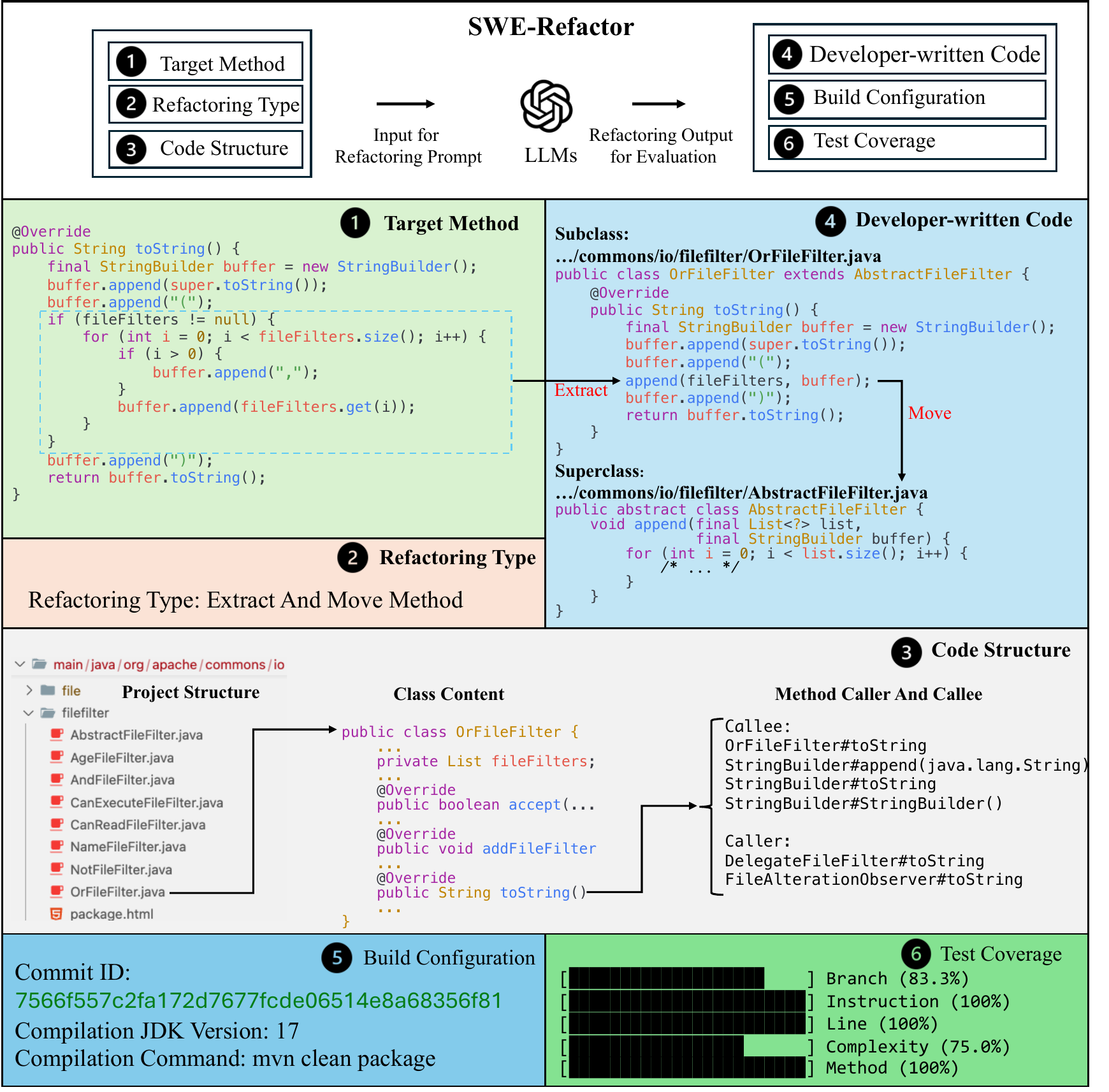} \centering
  \vspace{-0.2cm}
  \caption{An overview of the data in \tool.}
  \label{fig:overview}
  \vspace{-1mm}
\end{figure*}

Figure \ref{fig:overview} shows a data sample of \tool. Each sample in \tool contains 6 components.

\blackcircled{1} \textbf{Target Method}: The original method code before refactoring.
\blackcircled{2} \textbf{Refactoring Type}: The specific refactoring operation applied to the target method. For example, the data sample in Figure~\ref{fig:overview} illustrates an \textit{Extract and Move Method} refactoring, where a block of code is first extracted into a separate method and then moved to a more appropriate class.
\blackcircled{3} 
\textbf{Repository and Code Structure}: Structural information of the target method at the repository, class, and method levels. Repository-level details include the overall project structure and the full paths to all source Java files in the repository. Class-level details include the source code of the entire class and hierarchy (i.e., parent and child relationships).
Method-level information includes method’s callers and callees.
\blackcircled{4} \textbf{Developer-Written Code}: The target method refactored by project developers, serving as a reference for evaluating the quality of LLM-generated refactored code.
\blackcircled{5} \textbf{Build Configuration}: Compilation-related information necessary for building the project after refactoring. This includes the commit ID, the compatible JDK version, and the specific build commands.
\blackcircled{6} \textbf{Test Coverage}: Coverage data showing how the target method is exercised by the test suite. Comparing coverage before and after refactoring helps verify whether the refactoring preserves the program’s functional behavior.

\subsection{Task and Verification Metrics}
\label{sub:metrics}
As illustrated in Figure~\ref{fig:overview}, \tool is designed to evaluate the performance of Large Language Models (LLMs) in real-world code refactoring. Given a target method, a specific refactoring type, and relevant repository and source code information, \tool helps assess how effectively LLMs can generate correct and human-like refactored code. To evaluate refactoring quality from multiple perspectives, we employ three evaluation metrics: compilation and test success, \astm, and \textit{CodeBLEU}.


\blackcircled{1} \textbf{Compilation and Test success (Functional Verification)}. 
\tool integrates the LLM-generated refactored code into the project, then compiles the project and runs its test suites. This step verifies the functional correctness, ensuring the generated refactored code does not break the build or introduce unexpected issues.

\blackcircled{2} \textbf{\astm (Refactoring Verification)}. 
While compilation and test success reflect functional correctness, they do not guarantee that the intended refactoring has been applied and may risk overfitting to the test suite.
Due to potential hallucination issues in LLMs~\citep{DBLP:journals/corr/abs-2311-05232}, they may generate code that passes tests but deviates from the intended refactoring. To address this, we use \rfm \citep{tsantalis2020refactoringminer}, an Abstract Syntax Tree (AST) and rule-based static code analysis tool for detecting Java code refactorings, to verify whether the LLM-generated code contains the intended refactoring and to ensure the code contains no other functionality changes. \rfm has excellent performance at identifying refactorings within complex and mixed-purpose commits, achieving an average precision of 99\% and recall of 94\% in detecting refactoring \citep{tsantalis2020refactoringminer}. We manually inspect 200 sampled instances and confirm that all of them are pure, behavior-preserving refactorings without unrelated changes.

\blackcircled{3} \textbf{CodeBLEU (Human-Likeness Verfication)}. Finally, even when the code is functional and the refactoring is correct, it may still differ in quality or readability from the refactored code written by a human developer. Therefore, we include \textit{CodeBLEU}~\citep{ren2020codebleu} to assess the human-likeness of the generated code. 
\textit{CodeBLEU} is a code-specific evaluation metric that compares the textual, structural, and semantic similarities between two code snippets. By considering multiple dimensions, 
it provides a more accurate assessment of how closely the generated code matches what a human developer would write. 


 \subsection{Automated Benchmark Construction Pipeline}
 \begin{figure*}
\includegraphics[width=0.9\textwidth]{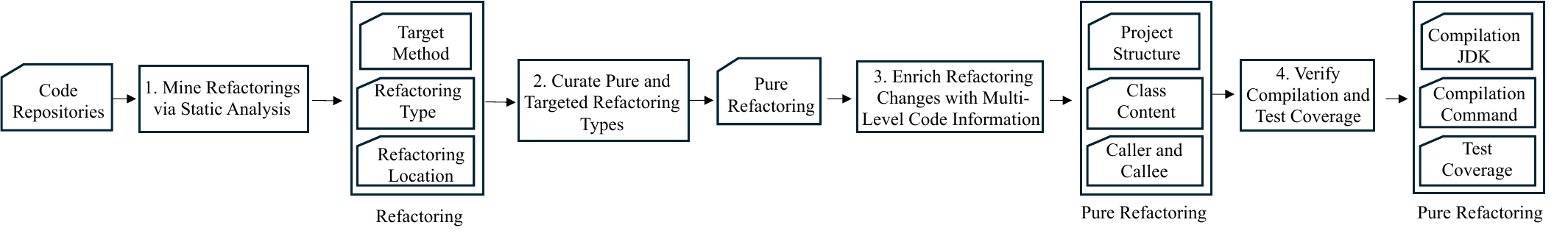} \centering
\vspace{-0cm}
  \caption{Our Automated Pipeline to Construct \tool.}
  \label{fig:pipeline}
  \vspace{-2mm}
\end{figure*}

Figure \ref{fig:pipeline} presents the automated pipeline of building \tool.  Unlike \textit{RefactorBench}
\citep{RefactorBench}, which synthesizes refactoring examples using LLMs, our dataset is built from real-world refactorings written by humans, identified through traditional static code and AST analysis. This design choice ensures the benchmark is free from LLM-induced hallucinations or bias. To construct \tool, we design a four-step automated pipeline: 

\textbf{Step 1: Mine Refactorings via Static Analysis}.
We leverage AST-based refactoring detection tools to extract commits that contain refactorings from GitHub repositories. \rfm is an AST- and rule-based tool that demonstrates high accuracy in refactoring detection. In addition to identifying refactoring types, we apply static code analysis to analyze the Java files. For each detected refactoring instance, we analyze the code and extract the detailed location information, including the commit hash, the affected Java files, and the specific line numbers within the file. This information is also stored in \tool as part of our released dataset. 
Based on this information, we further build the ASTs of the modified Java files. Then, we traverse the ASTs to extract Method Level and Class Level information for the refactoring instance, including the source code before and after the developer's refactoring changes, and the method and class signatures.

\textbf{Step 2: Curate Pure and Targeted Refactoring Types.} After extracting all commits containing refactorings, we use AST-based pure refactoring detection tools to curate high-quality instances by filtering out impure changes (e.g., bug fixes) and retaining only the six refactoring types studied in this work. \puc~\citep{nouri2023puritychecker} extends \rfm with specialized AST analysis to identify pure method-level refactorings, with an average precision of 95\% and recall of 88\%. It starts by identifying refactorings in a commit and comparing the code before and after the refactoring. During this process, \puc analyzes how original statements are changed—specifically, which statements were moved, modified, or replaced as part of the refactoring. It then checks whether these changes follow predefined purity rules. 


\textbf{Step 3: Enrich Refactoring Changes with Multi-Level Code Information.} \rfm analyzes refactorings within individual Java files and does not support cross-file analysis. Hence, we further use the Eclipse Java Development Tools (Eclipse JDT)~\citep{eclipseJDT} to extract structural information at the repository, class, and method levels. Eclipse JDT is a static analysis tool that provides access to the ASTs and type bindings of Java projects.
For each refactoring instance, we identify the modified Java files and collect additional source files within the same software package. We implement static analysis tools to analyze these files and construct ASTs with resolved types and method references. By traversing the ASTs, we extract the repository structure, the source code of the entire class and its hierarchy, and caller-callee relationships.

\textbf{Step 4: Verify Compilation and Test Coverage.} For each refactoring, we develop a script to compile the project and verify its correctness. To determine the appropriate JDK version, we attempt compilation using multiple JDKs. We then execute the test suite with JaCoCo \citep{jacoco} to collect code coverage information and exclude commits where the refactored code is not exercised by any test. Finally, we verify the existence of target classes involved in \textit{Move Method}, \textit{Extract and Move Method}, and \textit{Move and Inline Method} refactorings. This step was necessary because the \textit{Move Method} operation may move a method to newly created classes, and it is difficult for LLMs to predict the newly created classes. 

\label{sub:pipeline}

\section{Experiment}
\label{sec:experiment}


We evaluate 9 popular LLMs on \tool, and analyze their effectiveness across different refactoring types, prompting strategies, and multi-agent workflows. They cover general LLMs (i.e., gpt-4o-mini-2024-07-18 \citep{DBLP:journals/corr/abs-2303-08774}, gpt-3.5-turbo-01-25 \citep{GPT3.5}, and DeepSeek-V3 \citep{DBLP:journals/corr/abs-2412-19437}) and Code LLMs (Qwen2.5 Coder-\{7b, 14b\} \citep{DBLP:journals/corr/abs-2409-12186}, DeepSeek Coder-\{6.7B, 16B\} \citep{DBLP:journals/corr/abs-2401-14196}, and CodeLLaMa-\{7B,13B\} \citep{DBLP:journals/corr/abs-2308-12950}). 
General LLMs are accessed via official APIs, while Code LLMs are deployed on a cluster with 4 NVIDIA A100 GPUs (40GB each).

\subsection{LLMs' Performance on \tool}
\label{sub:overall_performance}
\begin{table*}[t]
\centering
\caption{Evaluation of 9 LLMs on \tool. The table presents the number of refactorings to perform, compile-and-test success rates, refactoring correctness verified by AST-Based refactoring detection tools (AST-Based RF Verification), and code similarity to human-written refactorings (\textit{Code BLEU}). \textbf{Successful Refactoring} refers to the number of refactorings that compile, pass tests, and are verified by AST-Based refactoring detection tools. We report the average \textit{Code BLEU} score and total counts for the other metrics.}
\label{tab:eval_result_1}
\scalebox{0.8}
{\setlength{\tabcolsep}{7mm}
\begin{tabular}{llllll} 
\hline
\multirow{2}{*}{\textbf{Model }} & \multirow{2}{*}{\textbf{Size }} & \textbf{Compile\&Test } & \textbf{AST-Based RF} & \textbf{Code} & \textbf{Successful } \\
 &  & \textbf{Success } & \textbf{Verification } & \textbf{BLEU} & \textbf{Refactoring } \\ 
\hline
gpt-4o-mini & N/A & 537 (48.86\%) & 636 (57.87\%) & 0.547 & 438 (39.85\%) \\
gpt-3.5-turbo & N/A & 199 (18.11\%) & 142 (12.92\%) & 0.536 & 82 (7.46\%) \\
DeepSeek-V3 & N/A & \textbf{554 (50.41\%)} & \textbf{674 (61.33\%)} & \textbf{0.584} & \textbf{457 (41.58\%)} \\
\hline
Qwen2.5 Coder & 14B & 22 (2.00\%) & 101 (9.19\%) & 0.428 & 7 (0.64\%) \\
Qwen2.5 Coder & 7B & 20 (1.82\%) & 142 (12.92\%) & 0.582 & 6 (0.55\%) \\
DeepSeek Coder & 16B & 23 (2.09\%) & 101 (9.19\%) & 0.549 & 3 (0.27\%) \\
DeepSeek Coder & 6.7B & 31 (2.82\%) & 70 (6.37\%) & 0.442 & 7 (0.64\%) \\
CodeLLaMa & 13B & 14 (1.27\%) & 15 (1.36\%) & 0.558 & 1 (0.09\%) \\
CodeLLaMa & 7B & 41 (3.73\%) & 48 (4.37\%) & 0.502 & 12 (1.10\%) \\
\hline
\end{tabular}
}
\end{table*}
We evaluate \numref pure refactorings from the \tool using the three metrics defined in Section~\ref{sub:pipeline}: Compilation and Test Success, \astm, and \textit{CodeBLEU}. A refactoring is considered successful if it passes both Compilation\&Tests and \astm. For consistency, we design a standardized prompt template containing four components: (1) a task description of the refactoring, (2) the target method, (3) repository-level context such as class source and caller–callee relations, and (4) a natural language instruction specifying the expected transformation. The detailed prompt template is provided in Appendix~\ref{appendix:prompt-templates}.
As shown in Table~\ref{tab:eval_result_1}, DeepSeek-V3 achieves the best overall performance with 457 successful refactorings (41.58\%), followed by GPT-4o-mini with 438 (39.85\%). General-purpose LLMs substantially outperform open-source code LLMs, reflecting their stronger capabilities in code understanding. Among the open-source models, CodeLLaMa-7B performs best with 12 successes (1.10\%), while the 13B variant performs worse, likely due to its Python-focused pre-training~\citep{chai2025mceval}, which highlights the importance of having a non-Python benchmark.

\subsection{Performance across Refactoring Types}
To better understand how LLMs perform on different kinds of refactorings, we analyze their effectiveness across the six refactoring types studied in \tool: three atomic types (\textit{Extract Method}, \textit{Move Method}, \textit{Inline Method}) and three compound types (\textit{Extract and Move Method}, \textit{Move and Inline Method}, and \textit{Move and Rename Method}). For each refactoring type, we compute the success rate based on Compilation and Test Success and \astm. This analysis helps reveal whether certain LLMs are more effective at atomic refactorings compared to compound ones, and whether some types pose more challenges for current models.
Table~\ref{tab:eval_refactoring_type} shows that DeepSeek-V3 achieves the strongest specialization on \textit{Extract Method} with 301 successes, while GPT-4o-mini exhibits broader generalization, particularly in cross-file tasks such as \textit{Move Method} (92) and \textit{Extract+Move} (33). Open-source models (Qwen2.5, DeepSeek Coder, and CodeLLaMa) succeed mainly only on a few \textit{Extract Method} instances. 

Overall, the table highlights a clear trend: \textbf{current LLMs remain effective on local atomic edits but perform poorly on cross-file and compound transformations}. These tasks thus represent critical benchmarks for advancing LLMs’ reasoning ability over structured software artifacts.




\begin{table*}[t]
\caption{Performance of LLMs across six refactoring types.
\textbf{EM} = Extract Method, \textbf{IM} = Inline Method, \textbf{MM} = Move Method, \textbf{RM} = Rename Method.
Values in parentheses indicate the total number of instances per refactoring type collected in the \tool.}
\begin{center}
\label{tab:eval_refactoring_type}

\scalebox{0.8}{\setlength{\tabcolsep}{5mm}
\begin{tabular}{llccccccc} 
\toprule
\multirow{2}{*}{\textbf{Model}} & \multirow{2}{*}{\textbf{Size}} & \textbf{Successful} & \textbf{EM} & \textbf{IM} & \textbf{MM} & \textbf{EM + MM} & \textbf{MM + RM} & \textbf{MM + IM} \\
  &  & \textbf{Refactoring}  & \textbf{(441)} & \textbf{(71)} & \textbf{(410)} & \textbf{(142)} & \textbf{(21)} & \textbf{(14)} \\ 
\hline 
gpt-4o-mini & N/A & 438 & 259 &\textbf{53} & \textbf{92} & \textbf{33} & \textbf{1} & 0 \\
gpt-3.5-turbo & N/A & 82 & 48 & 9 & 23 & 2 & 0 & 0 \\
DeepSeek-V3 & N/A & \textbf{457} & \textbf{301} & 50 & 76 & 30 & 0 & 0 \\
Qwen2.5 Coder & 14B & 7 & 2 & 5 & 0 & 0 & 0 & 0 \\
Qwen2.5 Coder & 7B & 6 & 5 & 1 & 0 & 0 & 0 & 0 \\
DeepSeek Coder & 16B & 3 & 1 & 1 & 0 & 1 & 0 & 0 \\
DeepSeek Coder & 6.7B & 7 & 6 & 1 & 0 & 0 & 0 & 0 \\
CodeLLaMa & 13B & 1 & 1 & 0 & 0 & 0 & 0 & 0 \\
CodeLLaMa & 7B & 12 & 12 & 0 & 0 & 0 & 0 & 0 \\
\bottomrule
\end{tabular}}
\end{center}
 \vspace{-3mm}
\end{table*}

\subsection{Impact of Context Augmentation and Multi-Agent Workflows}

\begin{figure}[t]
  \centering
  \includegraphics[width=\columnwidth]{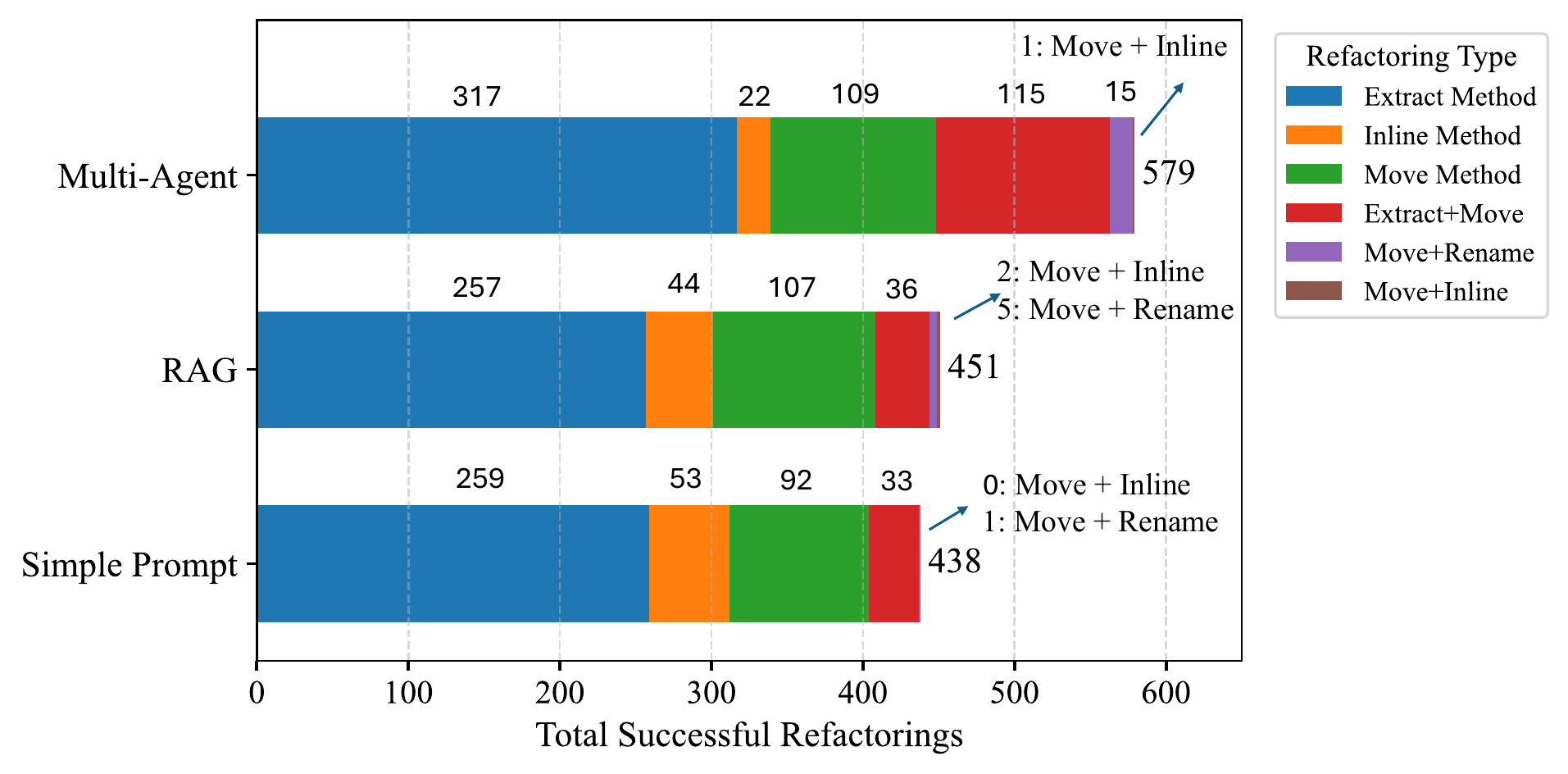}
  \vspace{-2mm}
  \caption{Comparison of successful refactorings.}
  \label{fig:eval_prompt}
  \vspace{-10mm}
\end{figure}


To examine the effect of context augmentation and multi-agent reasoning, we extend beyond simple prompting on \tool using two techniques. We apply Retrieval-Augmented Generation (RAG) to provide additional context via retrieved refactoring examples, and a multi-agent workflow that iteratively refines the outputs. We evaluate both techniques using \texttt{gpt-4o-mini}, chosen for its strong performance on complex refactorings and tool support.

RAG provides more context to LLMs through relevant few-shot examples, aiming to improve the accuracy and relevance of the generated code \citep{DBLP:journals/corr/abs-2410-09662, shirafuji2023refactoring}. Our RAG implementation uses a retrieval database of 905 pure refactoring instances drawn from the \textit{Refactoring Oracle Dataset} \citep{tsantalis2020refactoringminer}, which has no overlap with the data in \tool (construction details in Appendix~\ref{appendix:rag-construction}). 
The multi-agent workflow strengthens the reasoning and validation abilities of LLMs \citep{DBLP:journals/corr/abs-2312-13010}. We define two roles: a \textit{Developer Agent}, which generates refactored code given context, and a \textit{Reviewer Agent}, which critiques the output and provides iterative feedback. This design enables multi-turn refinement while mitigating common reasoning failures (Appendix~\ref{appendix:multi-agent}).

As shown in Figure~\ref{fig:eval_prompt}, the Multi-Agent strategy achieves the highest overall success (579 refactorings), outperforming RAG (451) and Simple Prompting (438). While all three perform similarly on \textit{Extract Method}, the Multi-Agent workflow shows clear advantages on more complex refactoring, completing 109 \textit{Move Method} and 115 \textit{Extract+Move} cases, far exceeding RAG (107, 36) and Simple Prompt (92, 33). These improvements likely stem from iterative reasoning and feedback between agents.

\subsection{Scalability to Agentic Scaffolding}
\label{sec:sota_models}
To further assess the limits of \tool, we extended our evaluation to an agentic scaffolding setup using OpenAI Codex (\texttt{GPT-5.1-Codex}) \cite{gpt-5-codex}.
 We conducted a stratified sample of 200 instances, constructed by considering the distribution of refactoring types and executable lines of code (ELOC). Specifically, we selected 100 instances with ELOC $\le$ 10 and 100 with ELOC $>$ 10. This resulted in 80 \textit{Extract}, 13 \textit{Inline}, 74 \textit{Move}, 26 \textit{Extract+Move}, 4 \textit{Move+Rename}, and 3 \textit{Move+Inline} instances. Codex was provided with full repository access and the same prompts used in our prior evaluation.

We find that \tool is challenging even for strong models. \textbf{GPT-5.1-Codex} solves 151/200 instances (75.5\%) overall, but the performance drops sharply from atomic to compound refactorings: 138/167 (82.6\%) for atomic versus only 13/33 (39.4\%) for compound. This large gap highlights the difficulty of compound refactorings in \tool and suggests that \textit{current LLM-based refactoring agents remain far from reliable on multi-step}, tightly constrained transformations. Most failures are caused by mismatched edits, where the model performs only part of the requested compound refactoring or produces an alternative transformation that does not satisfy the specified operation.

\section{Discussion}
\label{sec:discussion}

\textbf{Error Taxonomy.} 
To analyze failure modes, we sampled 50 refactorings for each of three representative settings: a small code LLM, a general LLM, and a multi-agent workflow. The small code LLM (i.e., CodeLLaMa-7B) failed on nearly all sampled cases, primarily because most outputs ignored the format requirements specified in the prompt, resulting in parsing errors. In contrast, the general LLM (i.e., GPT-4o-mini) was more reliable in following instructions but still showed weaknesses in handling code dependencies and repository-level information. Its major failures included syntax-level errors (e.g., undefined variables and parameter type mismatches) and semantic errors such as moving methods into non-existent files. The multi-agent workflow (using GPT-4o-mini) succeeded in most cases, though its remaining failures often reflected overfitting to the test cases. For example, generating empty methods that passed compilation and testing but failed \astm. 
The observed error patterns highlight the distinct strengths and weaknesses of different LLMs, RAG, and the multi-agent workflow. The results also show that \tool can assess LLM robustness at multiple levels, from following basic schema in small models to performing repository-level reasoning in multi-agent systems. 

\textbf{Limitations.} 
\tool has three main limitations. First, it focuses only on Java projects. While this limits language diversity, it enables reliable extraction using mature Java-based code analysis tools such as \rfm~\cite{tsantalis2020refactoringminer}, RefDiff~\cite{RefDiff2.0:2021}, and PMD~\cite{pmd}, and provides a valuable complement to existing Python-centric benchmarks. We plan to extend to other languages to support multi-language evaluation. Second, \tool currently targets method-level refactorings due to their high prevalence in real-world projects~\cite{kim2014empirical,Negara:2013}. Higher-level refactorings such as those at the class level are less frequent and often entangled with non-refactoring changes such as bug fixes~\cite{dipenta2020relationshiprefactoringactionsbugs}, which makes extraction more challenging. We aim to include a broader range of refactoring types in the future. Third, although \tool includes \numref pure refactorings from \numproj projects, making it one of the largest benchmarks of its kind, the scale is still limited for comprehensive evaluation or fine-tuning of LLMs. We plan to continue expanding the dataset to improve coverage and diversity.

\section{Conclusion}
\label{sec:conclusion}

In this work, we present \tool, a new benchmark specifically designed to evaluate the capabilities of LLMs in code refactoring. \tool features \numref pure, real-world refactorings extracted from \numproj diverse Java projects, covering both atomic and compound refactoring types. It ensures high data quality through automated filtering, compilation, and test verification, and includes rich repository-level information to support realistic and comprehensive evaluation. 
Evaluating 9 widely used LLMs shows that \tool remains challenging, especially for compound refactorings. For example, \textbf{GPT-5.1-Codex} achieves 39.4\% success on compound instances. We publicly release all data and results to support future research in LLM-based code refactoring.


\bibliography{example_paper}
\bibliographystyle{icml2026}

\newpage
\appendix
\onecolumn
\newpage
\appendix

\section*{Appendix}
\section*{Table of Contents}
\begin{itemize}

  \item \textbf{Appendix \ref{appendix:dataset-hosting}:} Dataset Hosting
  \item \textbf{Appendix \ref{appendix:llm_use}:} Use of Large Language Models (LLMs)
  \item \textbf{Appendix \ref{appendix:refactoring-type}:} Refactoring Type Definitions
  \item \textbf{Appendix \ref{appendix:dataset}:} Project Selection and Refactoring Distribution
  \item \textbf{Appendix \ref{appendix:prompt-templates}:} Prompt Templates for Different Refactoring Types
  \item \textbf{Appendix \ref{appendix:rag-construction}:} 
  RAG Construction for Refactoring Retrieval
  \item \textbf{Appendix \ref{appendix:multi-agent}:} Workflow For Multi-Agent
\end{itemize}

\section{Dataset Hosting}
\label{appendix:dataset-hosting}
Our \tool benchmark and experimental results (e.g., code, prompts, and LLM predictions) are available on the following platform:
\begin{itemize}
    \item \textbf{To be released}: 
\end{itemize}

\section{Use of Large Language Models (LLMs)}
\label{appendix:llm_use}
Large Language Models (LLMs) were used only to polish the writing. They were not involved in the research design, analysis, or conclusions.

\section{Refactoring Type Definitions}
\label{appendix:refactoring-type}

We define the refactoring types evaluated in this study based on widely accepted descriptions from Fowler’s Refactoring Catalog~\citep{DBLP:books/daglib/0019908} and \rfm~\citep{tsantalis2020refactoringminer}. These definitions serve as the foundation for identifying and categorizing both basic and compound refactorings in our benchmark.

\begin{itemize}
    \item \textbf{Extract Method.} 
    A code fragment is extracted from an existing method and placed into a newly created method. The original fragment is replaced with a method call. This improves readability, modularity, and reuse, especially when the original method becomes long or performs multiple responsibilities.

    \item \textbf{Move Method.}
    A method is relocated from one class to another, usually when it relies more on the data of the target class. This improves cohesion and reduces coupling between classes.

    \item \textbf{Inline Method.}
    A method is removed by replacing its invocations with its body. This is typically done when the method is too simple, no longer adds meaningful abstraction, or is used only once.

    \item \textbf{Extract and Move Method.}
    A compound refactoring where a code fragment is first extracted into a new method, and the resulting method is then moved to another class (often a superclass). This is useful when the extracted logic is generalizable or better fits in a shared parent class.

    \item \textbf{Move and Rename Method.}
    A method is moved to a different class and renamed during the process. The renaming helps to align the method name with its new context or to resolve naming conflicts.

    \item \textbf{Move and Inline Method.}
    A method is first moved to a new class and then inlined at all its call sites. This effectively eliminates the method definition while relocating its logic, typically used when the method becomes redundant after reorganization.
    
    \item \textbf{Extract Variable.}  
    Extracts part of an expression or a literal value into a new local variable. This improves readability and allows reuse of the extracted value. It is often applied to clarify complex expressions or remove duplication.

    \item \textbf{Rename Method.}  
    Changes the name of a method to better reflect its purpose or conform to naming conventions. This improves code readability and maintainability. All call sites must be updated accordingly.

    \item \textbf{Move Class.}  
    Relocates a class from one package or module to another. This helps improve package organization and reduce module dependencies. All references and imports must be updated.

    \item \textbf{Rename Class.}  
    Changes the name of a class to better reflect its role or to align with naming standards. This refactoring improves clarity and consistency. The renaming may also require updating file names and documentation.
\end{itemize}

\section{Project Selection and Refactoring Distribution}
\label{appendix:dataset}

We selected 18 Java projects previously used in change history tracking studies~\citep{DBLP:conf/icse/GrundCBHH21, 10.1145/3540250.3549079, Hasan:TSE:2024:CodeTracker2.0} based on three key criteria. First, the projects span diverse application domains, offering broad coverage of real-world software development practices. Second, each project has a rich development history, with over 2,000 commits, increasing the likelihood of discovering meaningful refactoring activities. Third, we ensured that the selected projects could be compiled and tested successfully after manual resolution of build issues, making it feasible to verify the correctness of the generated refactorings.

Table \ref{tab:proj_info} presents the selected Java projects along with the number of extracted pure refactorings for each project.

\begin{table}[t]
    \centering
    \caption{Overview of Java projects used in the construction of \tool.}
    \label{tab:proj_info}
    \renewcommand{\arraystretch}{1.25}
    \rowcolors{2}{gray!10}{white}
    \begin{tabular}{lrrr}
        \rowcolor{gray!20}
        \textbf{Project} & \textbf{\# Stars} & \textbf{\# Commits} & \textbf{\# Pure Refactorings} \\
        \toprule
        checkstyle         & 8,462  & 14,606  & 91  \\
        pmd                & 4,988  & 29,117  & 125 \\
        commons-lang       & 2,776  & 8,404   & 59  \\
        hibernate-search   & 512    & 15,716  & 89  \\
        junit4             & 8,529  & 2,513   & 18  \\
        commons-io         & 1,020  & 5,455   & 93  \\
        javaparser         & 5,682  & 9,607   & 56  \\
        junit5             & 6,523  & 8,990   & 105 \\
        hibernate-orm      & 6,091  & 20,638  & 63  \\
        mockito            & 15,032 & 6,236   & 4   \\
        gson	&	24080	&	2135	&	21	\\
guava	&	51140	&	7068	&	300	\\
jadx	&	45589	&	2512	&	18	\\
zxing	&	33605	&	3832	&	21	\\
shiro	&	4402	&	4222	&	2	\\
shenyu	&	8663	&	3680	&	22	\\
shardingsphere-elasticjob	&	8211	&	2473	&	3	\\
hertzbeat	&	6665	&	2632	&	9	\\
        \midrule
        \textbf{Total}     & \textbf{241,970} & \textbf{149,836} & \textbf{1099} \\
        \bottomrule
    \end{tabular}
\end{table}

\section{Prompt Templates for Different Refactoring Types}
\label{appendix:prompt-templates}
\begin{itemize}
    \item \textbf{Prompt Template for Extract Method, Inline Method Refactoring.} 
   
\begin{tcolorbox}[promptbox]
{\footnotesize
\begin{verbatim}
Task: 
You are an expert software engineer. You are given a code to be 
refactored. The objective is to refactor this code by performing 
given refactoring operation. This refactoring will improve code 
readability, maintainability, and modularity.
Code to be Refactored:
{code_to_refactor}
Class content:
{class_content}
Refactoring Operation:
{refactoring_operation}
Call Relationship:
{call_relationship}
Instructions:
1. Analyze the provided code and class content, apply relevant 
   refactoring operation to the code to be refactored.
2. If refactoring is performed, output the refactored_method_code 
   in the following format:
##########################
refactored_method_code
##########################
\end{verbatim}
}\end{tcolorbox}

    \item \textbf{Prompt Template for Move Method, Move And Rename Method Refactoring.}
   
\begin{tcolorbox}[promptbox]
{\footnotesize
\begin{verbatim}
Task: 
You are an expert software engineer. You are given a code to be 
refactored. The objective is to refactor this code by performing 
given refactoring operation. This refactoring will improve code 
readability, maintainability, and modularity.
Code to be Refactored:
{code_to_refactor}
Class content:
{class_content}
Refactoring Operation:
{refactoring_operation}
Call Relationship:
{call_relationship}
Project Structure:
{project_structure}
Instructions:
1. Analyze the provided code, class content, and project 
structure, apply move method refactoring to the code to be 
refactored, output the target file path, moved class code, 
and refactored method code. Need to move to an existing 
java file
The moved method code should be updated to the public 
static method. The refactored method code should use the 
moved class to call the moved method.
The target file path should be the path of the existing class
where the method is moved to.
2. If refactoring is performed, output the target file path, 
moved class code, and refactored method code in the following 
format:
##########################
target_file_path
##########################
moved_class_code
##########################
refactored_method_code
##########################
\end{verbatim}
}\end{tcolorbox}
\item \textbf{Prompt Template for Move And Inline Method Refactoring.}

\begin{tcolorbox}[promptbox]
{\footnotesize
\begin{verbatim}
Task: 
You are an expert software engineer. You are given a code to be 
refactored. The objective is to refactor this code by performing 
given refactoring operation. This refactoring will improve code 
readability, maintainability, and modularity.
Code to be Refactored: {code_to_refactor}
Class content: {class_content}
Refactoring Operation: {refactoring_operation}
Call Relationship: {call_relationship}
Project Structure: {project_structure}
Instructions:
1. Analyze the provided code, class content, and project 
structure, apply relevant refactoring operation to the 
code to be refactored, output the target file path.
2. If refactoring is performed, output the refactored class code 
in the following format:
##########################
target_file_path
##########################
refactored_class_code
##########################
\end{verbatim}

}\end{tcolorbox}
    \item \textbf{Prompt Template for Extract And Move Method Refactoring.}
\begin{tcolorbox}[promptbox]
\footnotesize{
\begin{verbatim}
Task: 
You are an expert software engineer. You are given a code to 
be refactored. The objective is to refactor this code by 
performing given refactoring operation. This refactoring will 
improve code readability, maintainability, and modularity.
Code to be Refactored: {code_to_refactor}
Class content: {class_content}
Refactoring Operation: {refactoring_operation}
Call Relationship: {call_relationship}
Project Structure: {project_structure}
File Path Before Refactoring:
{file_path_before_refactoring}
Instructions:
1. Analyze the provided code, class content, and project 
structure, apply relevant refactoring operation to the code 
to be refactored, and you need move the 
extracted method to another existing java file, output the 
target file path, extracted method code, refactored method code 
after refactoring.
The extracted method code should be the public static method.
The refactored method code should use the moved class to call the 
extracted method.
The target file path should be the path of the existing class 
where the method is moved to.
2. If refactoring is performed, output the refactored class code 
in the following format:
##########################
target_file_path
##########################
extracted_method_code
##########################
refactored_method_code
##########################
\end{verbatim}
}\end{tcolorbox}

\end{itemize}
\section{RAG Construction for Refactoring Retrieval}
\label{appendix:rag-construction}
\begin{figure*}
\includegraphics[width=0.9\textwidth]{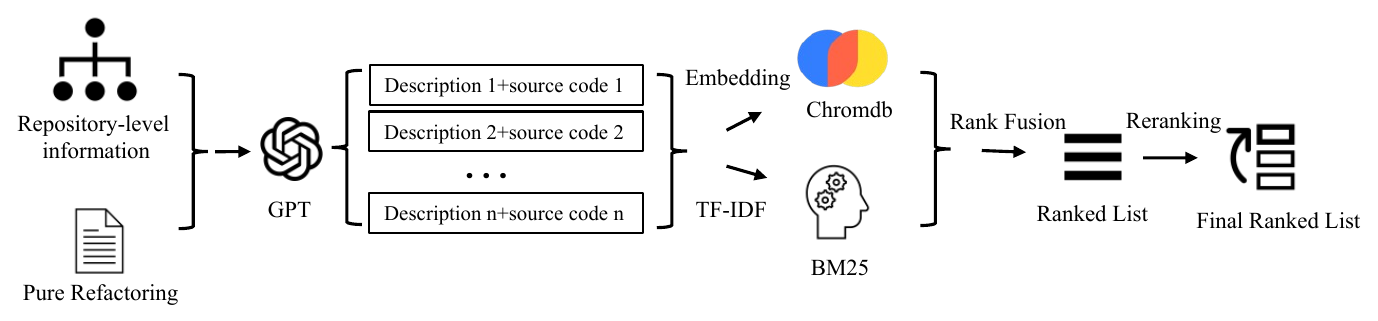} \centering
  \caption{RAG Construction and Retrieval Pipeline.}
  \label{fig:rag}
  \vspace{-3mm}
\end{figure*}

To support more accurate LLM-based code refactoring, we design a retrieval-augmented generation (RAG) pipeline. As shown in Figure~\ref{fig:rag}, it consists of four main steps: preparing the inputs, generating descriptions, retrieving relevant examples using both text and embedding similarity, and merging the results to find the most suitable matches.

\subsection*{Step 1: Preparing Inputs from Refactoring Commits}

We apply our pipeline (Section~\ref{sub:pipeline}) to the \textit{Refactoring Oracle Dataset}~\citep{tsantalis2020refactoringminer}, which contains over 12,000 refactorings collected from 547 commits across 188 open-source Java projects. This dataset has been widely used to evaluate refactoring detection tools and covers diverse projects and refactoring types. Using our pipeline, we extract a set of 905 pure method-level refactorings from this dataset. To save time, we do not perform compilation or test verification on these examples, as they are intended to illustrate refactoring strategies for retrieval rather than for correctness evaluation.

For each refactoring, we also collect repository-level information such as the file path, class definition, method signature, and the method’s direct callers and callees. These elements form the foundation of our retrieval database.

\subsection*{Step 2: Generating Descriptions of Refactoring Examples}

For each example, we use \texttt{gpt-4o-mini-0125} to generate a short natural language description that summarizes the method's functionality and surrounding structural information. The model takes as input the method before refactoring, its enclosing class, and the bodies of its direct callers and callees. These descriptions help guide retrieval by expressing the purpose and behavior of the method in a form that complements its code.

We use the following prompt template:

\begin{verbatim}
{Method Code}
{Caller/Callee Code}
{Class Code}
Please give a short, succinct description to situate this
code within the class.
\end{verbatim}

Here, \texttt{\{Method Code\}} is the code to be refactored, \texttt{\{Caller/Callee Code\}} includes the full bodies of its direct callers and callees, and \texttt{\{Class Code\}} provides the signature and body of the class containing the method.

\subsection*{Step 3: Constructing a Searchable Database of Refactoring Examples}
To support downstream retrieval, we construct a database of refactoring examples, where each entry includes both the code and its generated description. We index the database using two complementary methods to support both lexical and semantic similarity.

For text-based indexing, we apply BM25~\citep{robertson2009probabilistic}, which ranks examples based on token overlap and structural similarity in the combined code and description.

For semantic indexing, we use \texttt{all-MiniLM-L6-v2}~\citep{reimers-2019-sentence-bert} to generate vector embeddings for each example. This enables similarity computation based on meaning, not just syntax.


\subsection*{Step 4: Merging and Reranking the Results}

When a new refactoring task is issued, both text-based and embedding-based retrieval models produce independent similarity-ranked lists based on the input query. To combine these results, we apply the Reciprocal Rank Fusion (RRF) algorithm~\citep{DBLP:conf/sigir/CormackCB09}, which merges the rankings by assigning higher scores to examples that appear near the top of either list.

To further improve ranking quality, we apply a reranking step that refines the similarity assessment between the query and the retrieved examples. This step helps prioritize examples that are both lexically and semantically aligned with the input.

Finally, we select the top 3 ranked examples to serve as few-shot prompts, guiding the LLM to generate accurate and structurally relevant refactored code.

\section{Workflow for Multi-Agent}
\label{appendix:multi-agent}

To examine how multi-agent LLM workflows perform in automated code refactoring, we design a flexible agent-based system and evaluate it using our benchmark, \tool. The workflow is composed of two core agents: a \textit{Developer Agent} and a \textit{Reviewer Agent}. These agents communicate and collaborate through iterative reasoning and feedback.

\subsection*{Developer Agent: Generation and Refinement}

The \textit{Developer Agent} is tasked with analyzing source code and generating refactored code. It has three main capabilities: \textit{Analyzing}, \textit{Programming}, and \textit{Enhancing}. To support these tasks, the agent can invoke a variety of utility methods, such as retrieving project structure, reading source files, obtaining class body, or getting callers and callees. These methods are implemented through command-line tools or APIs from static analysis frameworks. After collecting the necessary information, the agent composes a prompt combining structural analysis and submits it to the LLMs to produce a refactored version of the target method. The agent can also iteratively improve its output by incorporating feedback received from the \textit{Reviewer Agent}.

\subsection*{Reviewer Agent: Evaluation and Feedback}

The \textit{Reviewer Agent} is responsible for assessing the quality of the generated refactoring. It performs this assessment by applying static analysis tools, including a refactoring detector (e.g., \rfm~\citep{tsantalis2020refactoringminer}) and a style checker (e.g., Checkstyle~\citep{checkstyle}) to detect code smells or violations of coding conventions. Based on this analysis, the \textit{Reviewer Agent} generates feedback indicating whether the refactoring is valid, and if not, what aspects should be improved. This feedback is then sent back to the \textit{Developer Agent} for further refinement.


\end{document}